\documentclass[prl,twocolumn]{revtex4}
\usepackage{graphicx}
\usepackage{amssymb,amsmath,bm}
\usepackage{enumerate}

\newcommand{\be}{\begin{eqnarray}}
\newcommand{\ee}{\end{eqnarray}}

\begin{document}

\title{Buoyancy driven turbulence and distributed chaos}

\author{A. Bershadskii}

\affiliation{
ICAR, P.O. Box 31155, Jerusalem 91000, Israel
}

\begin{abstract}

It is shown, using results of recent direct numerical simulations, laboratory experiments and atmospheric measurements, that buoyancy driven turbulence exhibits a broad diversity of the types of distributed chaos with its stretched exponential spectrum $\exp(-k/k_{\beta})^{\beta}$. The distributed chaos with $\beta = 1/3$  (determined by the helicity correlation integral) is the most common feature of the stably stratified turbulence (due to the strong helical waves presence). These waves mostly dominate spectral properties of the vertical component of velocity field, while the horizontal component is dominated by the diffusive processes both for the weak and strong stable stratification ($\beta =2/3$). For the last case influence of the low boundary can overcome the wave effects and result in $\beta =1/2$ for the vertical component of the velocity field (the spontaneous breaking of the space translational symmetry - homogeneity). For the unstably stratified turbulence in the Rayleigh-Taylor mixing zone the diffusive processes ($\beta =2/3$) are the most common dominating processes in the anisotropic chaotic mixing of the two fluids under buoyancy forces. The distributed chaos in Rayleigh-B\'{e}nard turbulent convection in an upright cell is determined by the strong confinement conditions. That is: the spontaneous breaking of the space translational symmetry (homogeneity) by the finite boundaries ($\beta = 1/2$) or by the non-perfect orientation of the cell along the buoyancy direction ($\beta =4/7$). In all types of turbulence appearance of an inertial range of scales results in deformation of the distributed chaos and $\beta =3/5$.

\end{abstract}

\maketitle

\section{Inroduction}

  For the simplest description of the stable and unstable buoyancy driven turbulence the Boussinesq approximation with an imposed linear temperature gradient is usually used
$$ 
\frac{ \partial{\mathbf u}}{\partial t} +{\mathbf u} \cdot \nabla {\mathbf u}  = -\frac{1}{\rho_{0}} \nabla P - N \theta  {\bf e_g} + \nu \nabla^2 {\mathbf u}+{\bf f}  \eqno{(1)}, 
$$ 
$$     
\frac {\partial \theta}{\partial t} +{\mathbf u} \cdot \nabla \theta  = s~N~ {\bf u} \cdot {\bf e_z} + D \nabla^2 \theta \eqno{(2)},
$$  
$$
 \nabla \cdot {\bf u} = 0 \eqno{(3)};
$$
where the buoyancy field $\theta$ is rescaled as a velocity.
  In the case of stable stratification the parameter $s=1$ in the unstable case $s=-1$, $ {\bf e_g}$ is an unit vector in the buoyancy direction (usually direction of the gravity acceleration ${\bf g}$), ${\bf e_z}$ is an unit vector in direction of the z-axis (usually, but not always - see below, ${\bf e_g} = {\bf e_z}$). In the case of stable stratification the parameter $N=\sqrt{-(g/\theta ) (d\bar \theta /dz)}$ is called the Brunt-V\"ais\"al\"a frequency, where 
$d{\bar \theta} /dz$ is the imposed linear temperature gradient (in this case $N$ is frequency of
the gravity waves). In the case of the Rayleigh-Taylor mixing zone (considered as an unstably stratified homogeneous
turbulence \cite{bur}) $N =  \sqrt{2g A (d{\bar c} /dz)}$, where $A$ is the Atwood number, $d{\bar c} /dz$ is the mean concentration z-gradient of heavy fluid and $\rho_{0} = (\rho_{heavy} + \rho_{light})/2$. 

  The case of the Rayleigh-B\'{e}nard convection can be also described by the Eqs. (1)-(3) if we replace original temperature $T$ by $\theta = T - \Delta T \cdot z/H$ (with corresponding pressure replacement in the case ${\bf e_g} = {\bf e_z}$, i.e. ideally horizontal Rayleigh-B\'{e}nard layer). Here $H$ is the layer height, $T_{low}$ is the low boundary temperature and $T_{low} + \Delta T$ is the upper boundary temperature.  The new temperature $\theta = 0$ along the low and upper boundaries (and it is rescaled as a velocity), $N=\sqrt{g\delta \Delta T/H}$ ($g$ is the gravity acceleration, $\delta$ is the thermal expansion coefficient), and $s = -1$. 
  
\section{Buoyancy driven distributed chaos}

{\bf A}. The Noether's theorem relates the space translational symmetry (homogeneity) to the momentum conservation \cite{ll2} and, therefore to the Birkhoff-Saffman integral 
$$
I_2 =\int  \langle {\bf u} \cdot  {\bf u'} \rangle d{\bf r} \eqno{(4)}
$$   
with ${\bf u'} ={\bf u} ({\bf x} + {\bf r},t) $ and ${\bf u} = {\bf u} ({\bf x},t)$. Namely, in homogeneous turbulence this integral is an invariant \cite{saf},\cite{dav1},\cite{dav2}. It is shown in Ref. \cite{b1} that this integral dominates distributed chaos in isotropic homogeneous turbulence resulting in the stretched exponential spectrum
$$
E (k) \propto \exp(-k/k_{\beta})^{\beta}  \eqno{(5)}
$$
with $\beta =3/4$. This value of $\beta$ was obtained in Ref. \cite{b1} using an asymptotic scaling approach to the group velocity $\upsilon (\kappa )$ of the waves driving the distributed chaos
$$
\upsilon (\kappa )\propto I_2^{1/2}~\kappa^{\alpha} \eqno{(6)}
$$
and $\alpha = 3/2$ from the dimensional considerations. General relation between $\alpha$ and $\beta$ according to the Ref. \cite{b1}
$$
\beta =\frac{2\alpha}{1+2\alpha}   \eqno{(7)}
$$ 
gives $\beta =3/4$ in this case.

   Let us consider a buoyancy generalization of the Birkhoff-Saffman integral
$$
I_b =   \int  \langle {\bf u} \cdot  {\bf u'} + s~ \theta~\theta' \rangle  d{\bf r}  \eqno{(8)}  
$$ 
(the buoyancy field $\theta$ is rescaled in the Eqs. (1)-(3) as a velocity). In the same manner as it is shown that the homogeneity results in the invariance of the Birkhoff-Saffman integral Eq. (4) for the ordinary turbulence (without buoyancy), it can be readily shown that the homogeneity results in invariance of the generalized  Birkhoff-Saffman integral for the turbulence described by the buoyancy Eqs. (1)-(3) with ${\bf f} = 0$ if ${\bf e_g} = {\bf e_z}$. Since the dimension of the integral $I_b$ Eq. (8) is the same as dimension of the integral $I_2$ Eq. (4) the value of $\beta$ also will be the same.\\

   {\bf B}. In a recent Ref. \cite{nrz} a spontaneous breaking of space translational symmetry (homogeneity) was studied for weak turbulence. For strong turbulence the spontaneous breaking was studied by the means of the distributed chaos in Ref. \cite{b2}. It is shown in the Ref. \cite{b2} that spontaneous breaking of the space translational symmetry (homogeneity), related to the finite boundary conditions, results in the distributed chaos dominated by vorticity correlation integral
$$
\gamma = \int_{V} \langle {\boldsymbol \omega} \cdot  {\boldsymbol \omega'} \rangle_{V}  d{\bf r} \eqno{(9)}. 
$$  
Substituting $\gamma$ into the scaling relation Eq. (6) instead of $I_2$ and using the dimensional considerations one obtains
$$
\upsilon (\kappa ) \propto |\gamma|^{1/2}~\kappa^{1/2} \eqno{(10)}
$$
and, consequently, $\beta =1/2$. In the same vein for the buoyancy driven turbulence we obtain distributed chaos dominated by the integral 
$$
\gamma_b = \int_{V} \langle {\boldsymbol \omega} \cdot  {\boldsymbol \omega'} + s Pr^{-1} \nabla \theta \cdot \nabla \theta' \rangle_{V}  d{\bf r} \eqno{(11)}. 
$$  
(where the Prandtl number $Pr= \nu/D$) for the spontaneous breaking of the space translational symmetry (homogeneity) by the finite boundary conditions. Since $\gamma_b$ has the same dimension
as $\gamma$ corresponding value of $\beta$ is the same 1/2.\\

\begin{figure}
\begin{center}
\includegraphics[width=8cm \vspace{-1cm}]{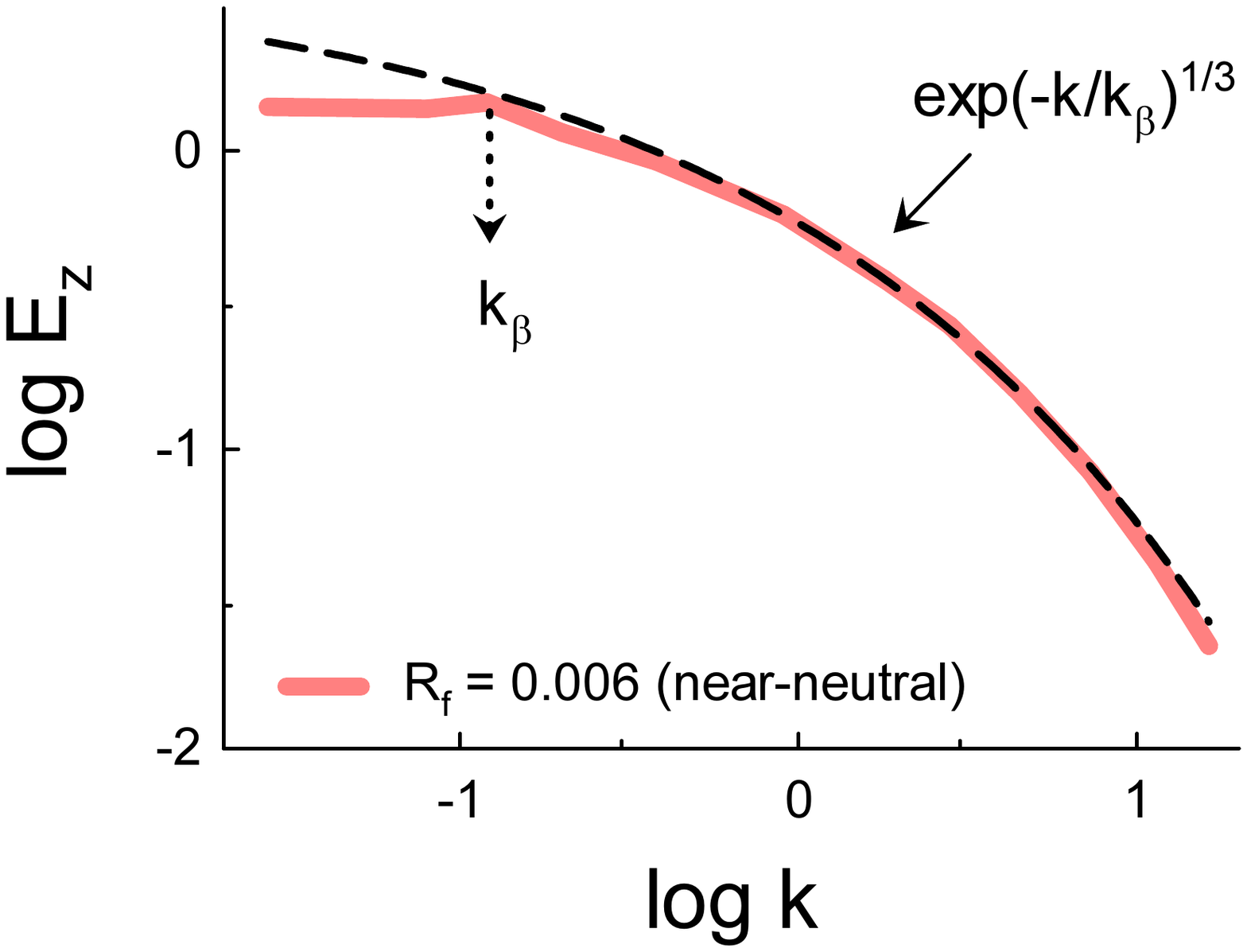}\vspace{-3.5cm}
\caption{\label{fig1} Power spectrum of vertical velocity for near-neutral stability conditions in the stable atmospheric surface layer over a lake. The dashed line indicates the stretched exponential spectral law Eq. (5) with $\beta = 1/3$. }
\end{center}
\end{figure}

\begin{figure}
\begin{center}
\includegraphics[width=8cm \vspace{-1.5cm}]{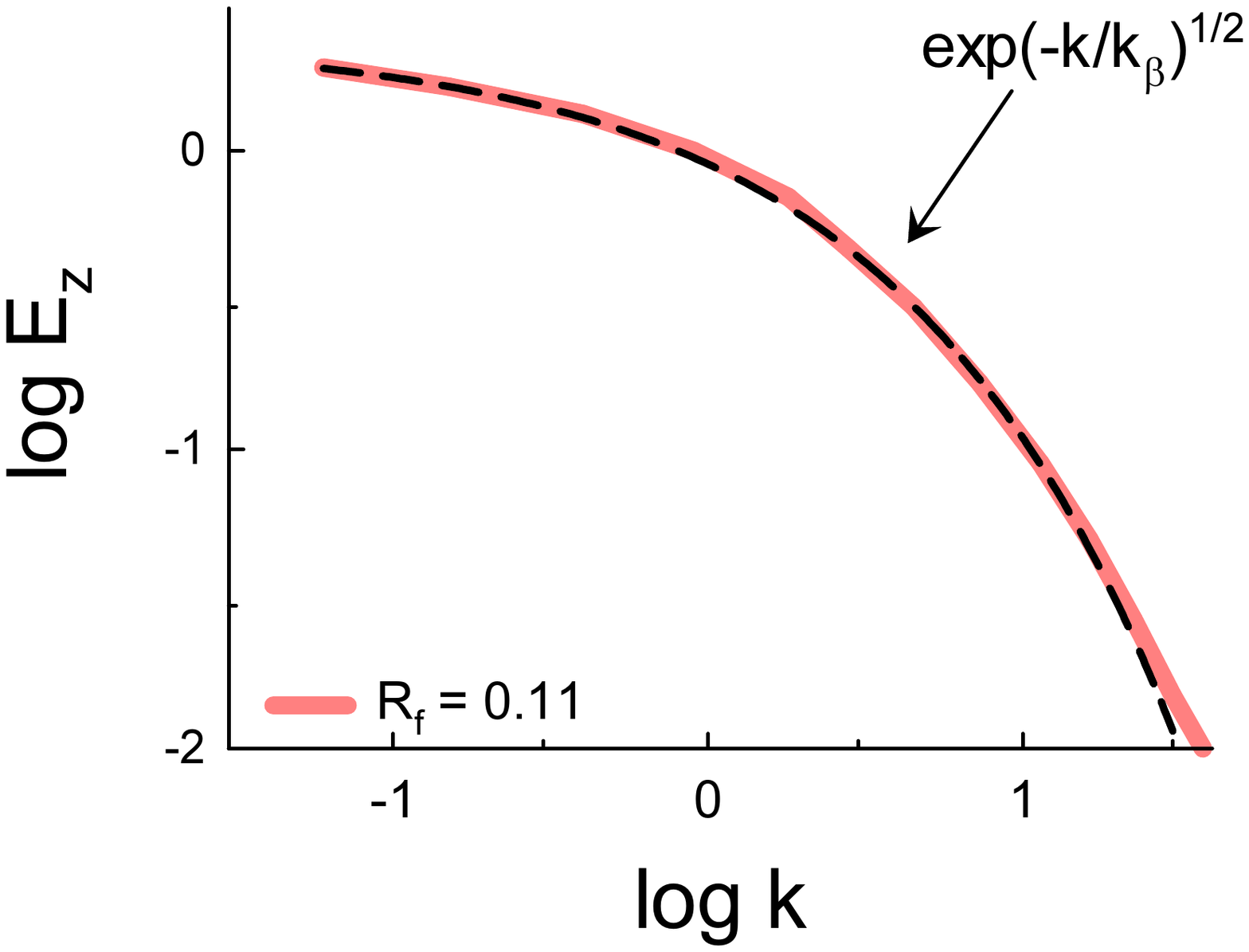}\vspace{-2.3cm}
\caption{\label{fig2} The same as in Fig.  1 but for rather stable situation. The dashed line indicates the stretched exponential spectral law Eq. (5) with $\beta = 1/2$. }. 
\end{center}
\end{figure}

  {\bf C}. An inertial range of scales can appear in the buoyancy driven turbulence at certain conditions \cite{my}. It is shown in Ref. \cite{b3} that appearance of inertial range of scales can also result in an (internal) spontaneous breaking of the space translational symmetry in the distributed chaos, and this distributed chaos is dominated by the energy correlation integral
$$
\mathcal{E}  = \int_V  \langle {\bf u}^2  \cdot  {\bf u'}^2  \rangle_V d{\bf r}  \eqno{(12)}
$$ 
Substituting the $\mathcal{E}$ into the scaling relation Eq. (6) instead of $I_2$ and using the dimensional considerations one obtains
$$
\upsilon (\kappa ) \propto ~\mathcal{E}^{1/4}~\kappa^{3/4} \eqno{(13)}
$$
and, consequently, $\beta =3/5$. For the generalized energy correlation integral
$$
\mathcal{E}_b  = \int_V  \langle ({\bf u}^2 + s ~\theta^2) \cdot  ({\bf u'}^2 + s ~\theta'^2)  \rangle_V d{\bf r}  \eqno{(14)}
$$ 
we obtain the same value of $\beta = 3/5$ for the buoyancy driven turbulence as well.\\

  {\bf D}. If ${\bf e_g} \neq {\bf e_z}$ (but still ${\bf f} =0$), then 
$$
\frac{dI_b}{dt} = -2\nu \gamma_b + G_b  \eqno{(15)}
$$
where the term $G_b \neq 0$ comes from the buoyancy terms $N \theta  {\bf e_g}$ and $sN{\bf u} \cdot {\bf e_z}$ (and from the pressure term in the case of the Rayleigh-B\'{e}nard convection) in the Eqs. (1),(2). It takes place, for instance, in the laboratory experiments with an upright Rayleigh-B\'{e}nard cell convection, due to non-ideal orientation of its 'vertical' z-axis along the buoyancy direction ${\bf g}$. This can results in a specific spontaneous breaking of the space translational symmetry (homogeneity). Substituting $|G_b|$ into Eq. (6) instead of $I_2$ and using the dimensional considerations one obtains
$$
\upsilon (\kappa ) \propto ~|G_b|^{1/3}~\kappa^{2/3} \eqno{(16)}
$$ 
and, consequently, $\beta =4/7$.\\
\begin{figure}
\begin{center}
\includegraphics[width=8cm \vspace{-1.4cm}]{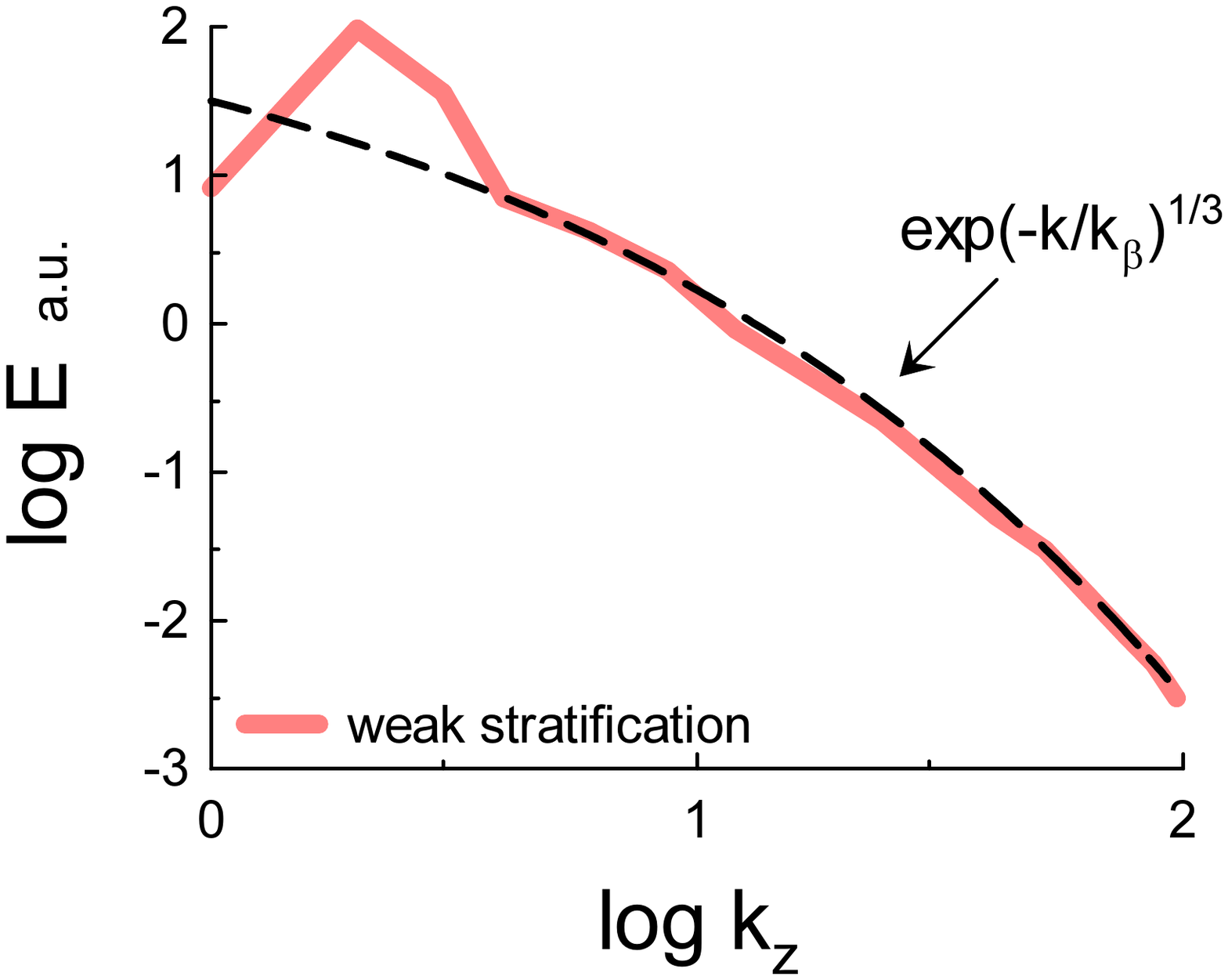}\vspace{-3.2cm}
\caption{\label{fig3} Vertical kinetic energy spectrum  for {\it weakly} stratified turbulence (DNS). The dashed line indicates the stretched exponential spectral law Eq. (5) with $\beta = 1/3$. }
\end{center}
\end{figure}  
 \begin{figure}
\begin{center}
\includegraphics[width=8cm \vspace{-1.2cm}]{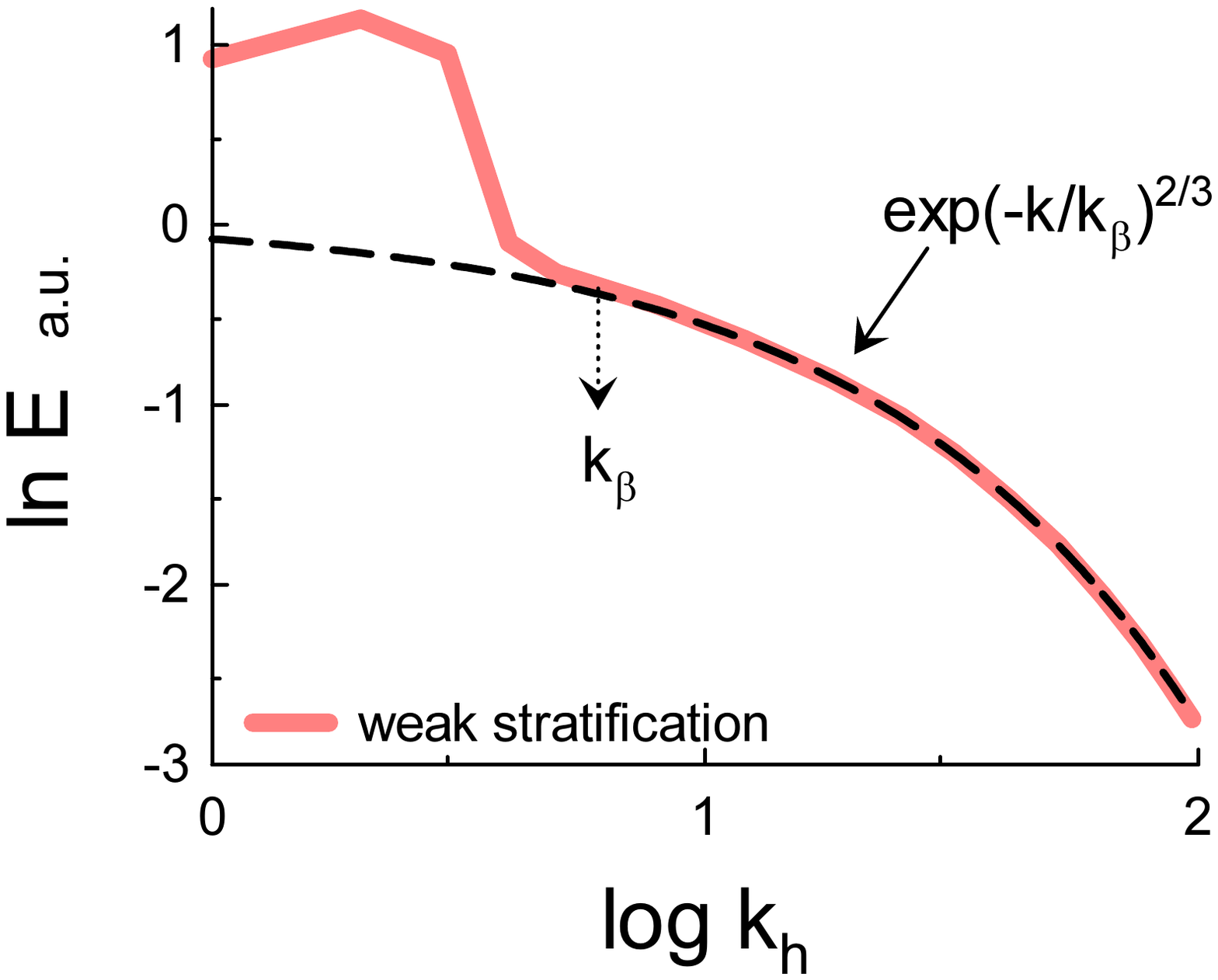}\vspace{-4cm}
\caption{\label{fig4} The same as in Fig. 3 but for horizontal kinetic energy. The dashed line indicates the stretched exponential spectral law Eq. (5) with $\beta = 2/3$.} 
\end{center}
\end{figure} 

\begin{figure}
\begin{center}
\includegraphics[width=8cm \vspace{-1.8cm}]{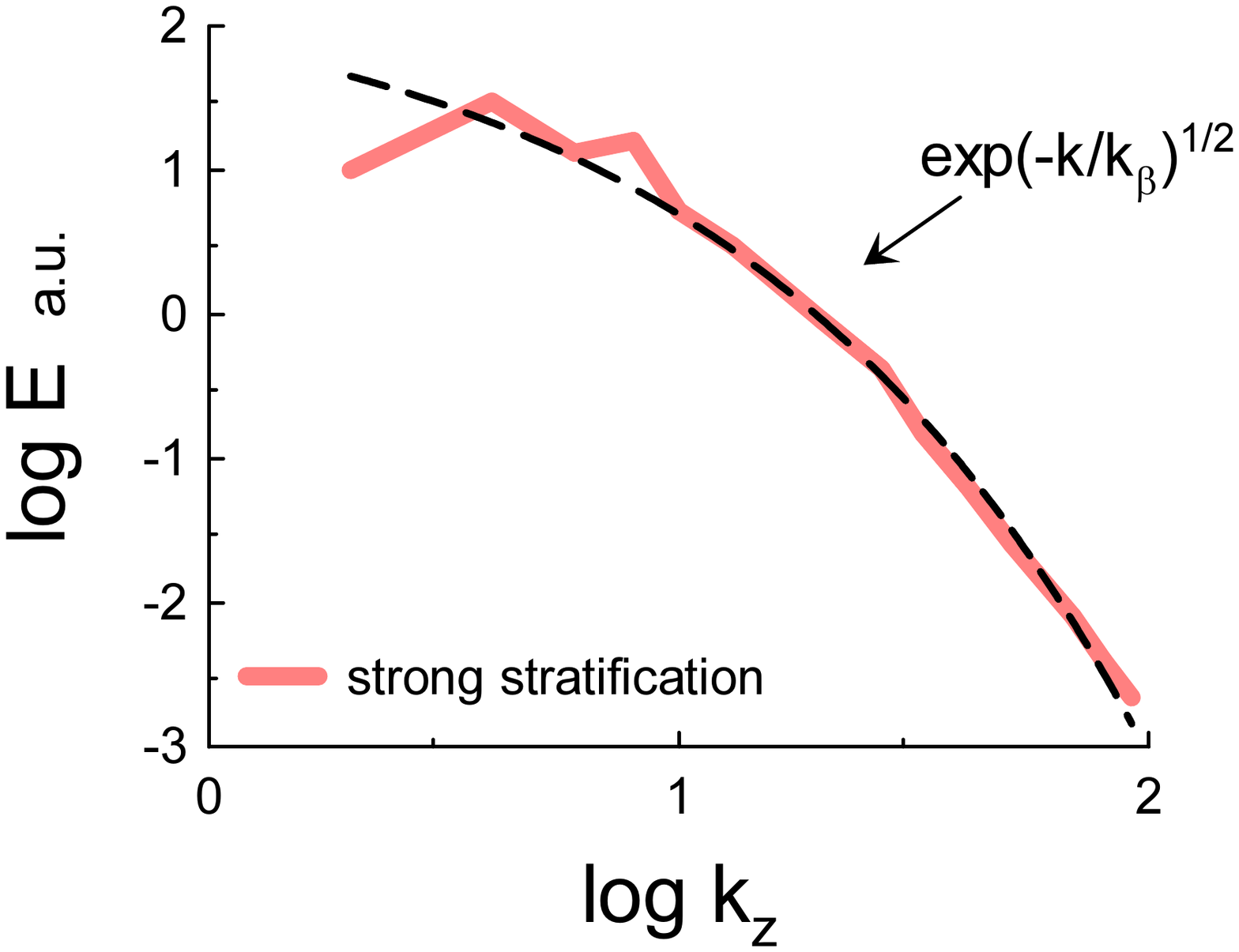}\vspace{-2.4cm}
\caption{\label{fig5} The same as in Fig. 3 but for {\it strongly} stratified turbulence. The dashed line indicates the stretched exponential spectral law Eq. (5) with $\beta = 1/2$.} 
\end{center}
\end{figure} 

  {\bf E}.  Till now we did not actually pay attention to anisotropy of the buoyancy driven turbulence (except the rather specific case {\bf D}). But this anisotropy can, under certain conditions, destroy all above considered cases related to the generalized momentum integral $I_b$ Eq. (8) (cf also the Ref. \cite{dav2}). In this case we have two possible dimensional parameters which can determines the distributed chaos: viscosity (mentioned as such in the Ref. \cite{b1}) and the helicity correlation integral
$$
I = \int_V \langle h~h'\rangle_V d{\bf r} \eqno{(17),}
$$
the Levich-Tsinober integral \cite{lt}, usually related to the helical waves \cite{l},\cite{b4}.

  In the first case substituting viscosity $\nu$ into Eq. (6) instead of $I_2$ one obtains from the dimensional considerations
$$
\upsilon (\kappa ) \propto ~\nu~\kappa \eqno{(18)}
$$
and, consequently, $\beta =2/3$ \cite{b1}.

  In the second case 
$$
\upsilon (\kappa ) \propto ~I^{1/4}~\kappa^{1/4} \eqno{(19)}
$$
and, consequently, $\beta =1/3$ \cite{b4}.\\

In the next three sections an application of the above described approach to different types of the buoyancy driving turbulence will be considered and then in the last section (Discussion) physical consequences of this consideration will be summarized. 

\section{Stably stratified turbulence}

  Let us start the data analysis from the atmospheric surface layer measurements. 
Figures 1 and 2 show (in the log-log scales) power spectra of vertical velocity measured in the stable atmospheric surface layer (over a lake) for the two stability regimes - from near-neutral to rather stable, respectively. The data were taken from Fig. 1 of the Ref. \cite{katul}  ($k$ is a wavenumber in the streamwise direction, normalized by $z$).  The dashed lines indicate the stretched exponential spectral law Eq. (5) with $\beta = 1/3$ (Section II, subsection E) for the  near-neutral and $\beta = 1/2$ (Section II, subsection B) for the rather stable conditions. The dotted arrow in Fig. 1 indicates tuning of the distributed chaos to the large-scale coherent structures (waves).

   The anisotropy is well demonstrated by the figures 3-5, which show (in the log-log scales) both vertical and horizontal kinetic energy spectra obtained in a recent direct numerical simulation (DNS) of stably stratified turbulence and reported in Ref. \cite{rem}. This DNS was performed in the presence of small-scale forcing applied to horizontal wavenumbers of the horizontal velocity components (with an approximately constant level of the turbulence kinetic
energy and without no mean velocity).
   
   Figure 3 shows vertical spectrum for {\it weakly} stratified turbulence and figure 4 shows corresponding horizontal spectrum. The dashed lines indicate the stretched exponential spectral law with $\beta = 1/3$ (Fig. 3) for the vertical spectrum and with $\beta = 2/3$ (Fig. 4) for the horizontal spectrum (Section II, subsection E). Figure 5 shows the vertical spectrum for {\it strongly} stratified turbulence. The dashed line indicates the stretched exponential spectral law  with $\beta = 1/2$ (Section II, subsection B). We did not show corresponding horizontal spectrum because it is rather similar (in its distributed chaos part) to that shown in Fig. 4.
   
   It is interesting to compare the results of the DNS with the results of the atmospheric measurements (Figs. 1 and 2). Because in this DNS the small-scale forcing was used it is also interesting to look to results of a DNS with {\it large}-scale forcing reported in the Ref. \cite{rmp}. Randomly generated 3D isotropic flows were used for the velocity forcing and the initial conditions in this DNS. The Reynold number $Re \simeq 25000$ and the Prandtl number $Pr = 1$. The data were taken from the Fig. 6a of the Ref. \cite{rmp}.
   
   Figure 6 shows total energy spectra ($k$ is isotropic wave number) for $N = 12$ (the Froude number $Fr = 0.03$). The dashed line indicates the stretched exponential spectral law with $\beta = 1/3$ (Section II, subsection E). Figure 7 shows the total energy spectra for $N = 4$ (the Froude number $Fr = 0.1$). The dashed line indicates the stretched exponential spectral law Eq. (5) with $\beta = 3/5$ (Section II, subsection C). An indication of the inertial range is shown by the straight solid line with the slope -5/3. The data were taken from the Fig. 6a of the Ref. \cite{rmp}.
   
\begin{figure}
\begin{center}
\includegraphics[width=8cm \vspace{-1.1cm}]{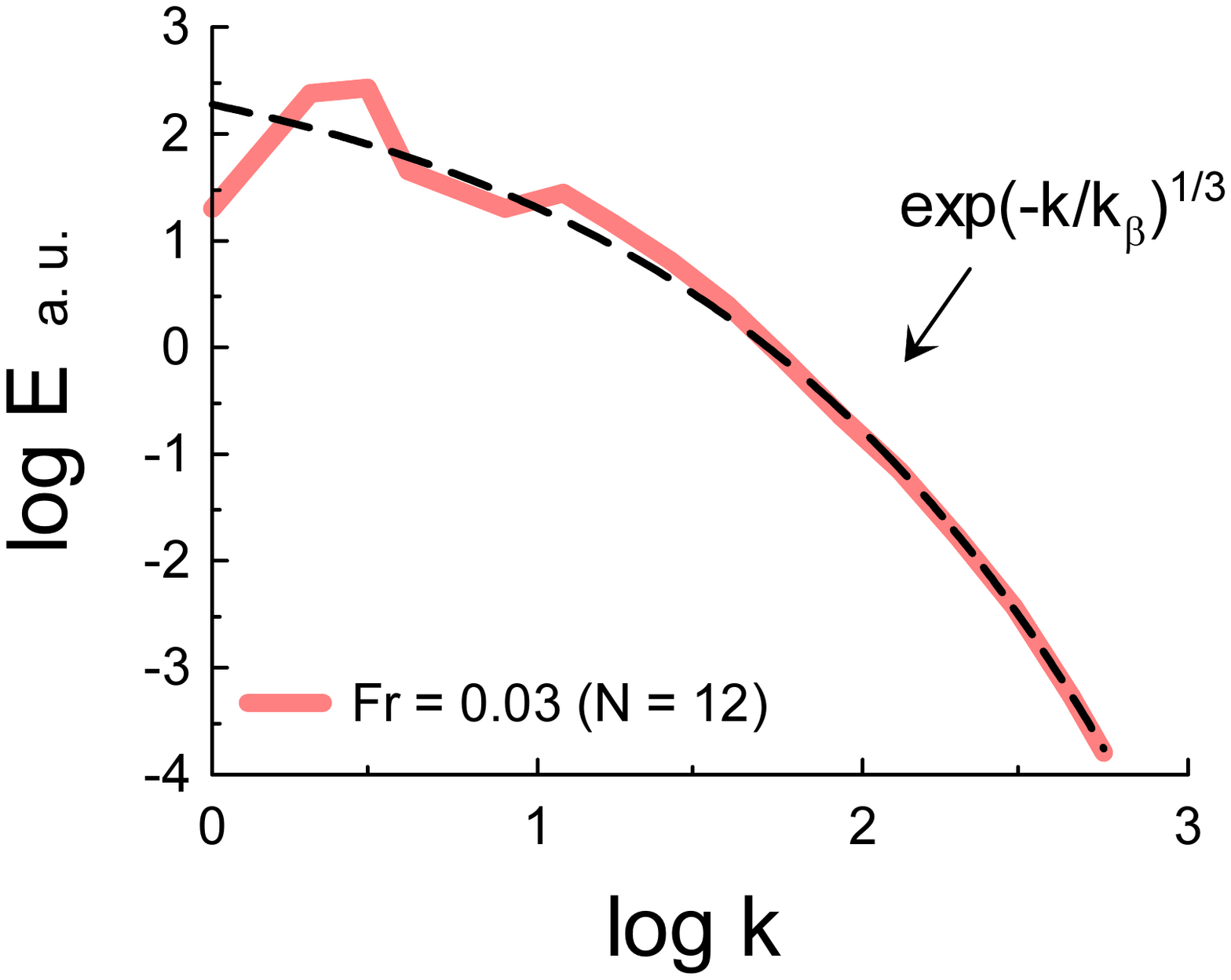}\vspace{-3.4cm}
\caption{\label{fig6} Total energy spectra ($k$ is isotropic wave number) for $N = 12$ ($Fr = 0.03$). The dashed line indicates the stretched exponential spectral law Eq. (5) with $\beta = 1/3$. }
\end{center}
\end{figure} 
\begin{figure}
\begin{center}
\includegraphics[width=8cm \vspace{-0.93cm}]{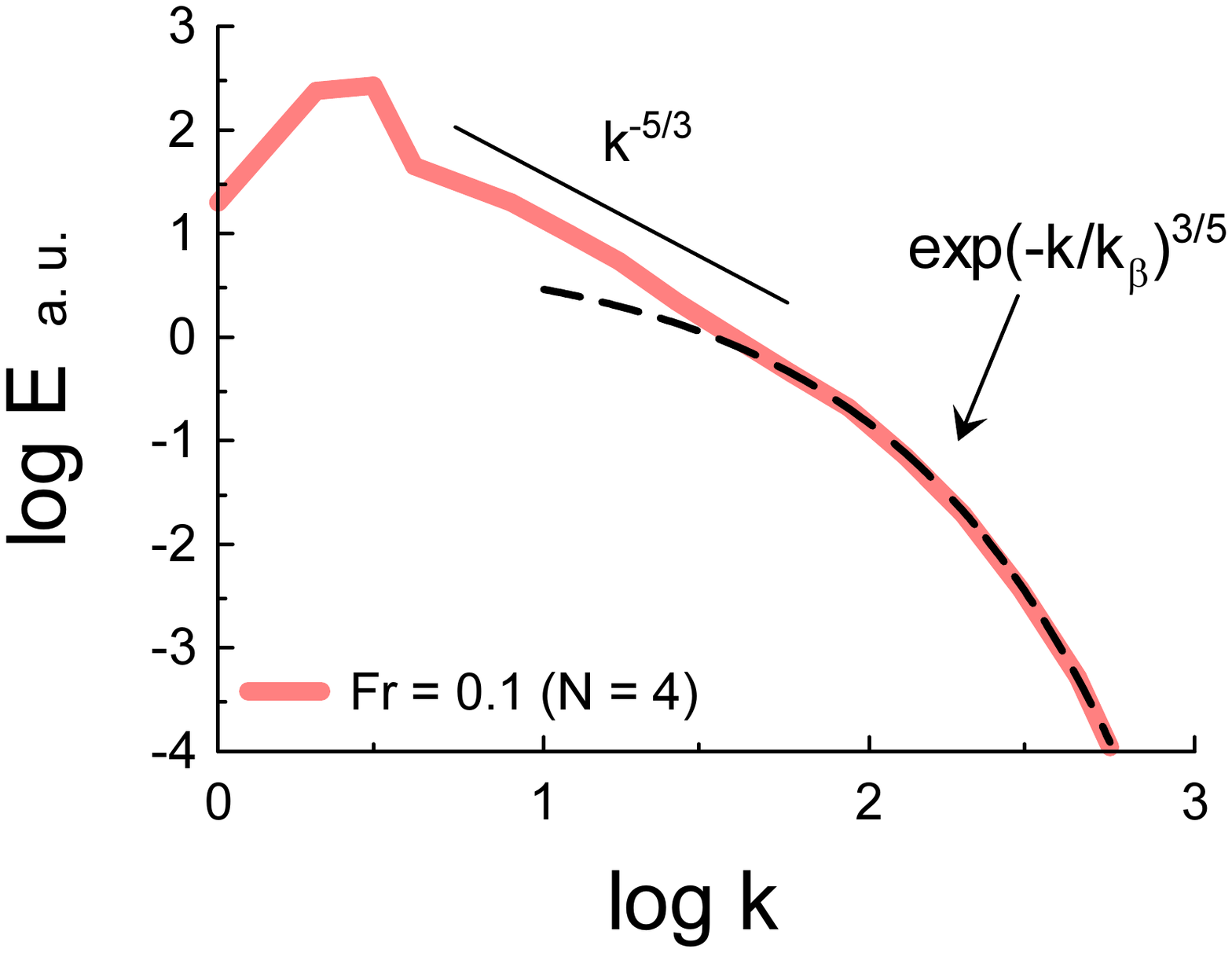}\vspace{-3.7cm}
\caption{\label{fig7} The same as in Fig. 6 but for $N = 4$ (the Froude number $Fr = 0.1$). The dashed line indicates the stretched exponential spectral law Eq. (5) with $\beta = 3/5$.  } 
\end{center}
\end{figure}
\begin{figure}
\begin{center}
\includegraphics[width=8cm \vspace{-1.1cm}]{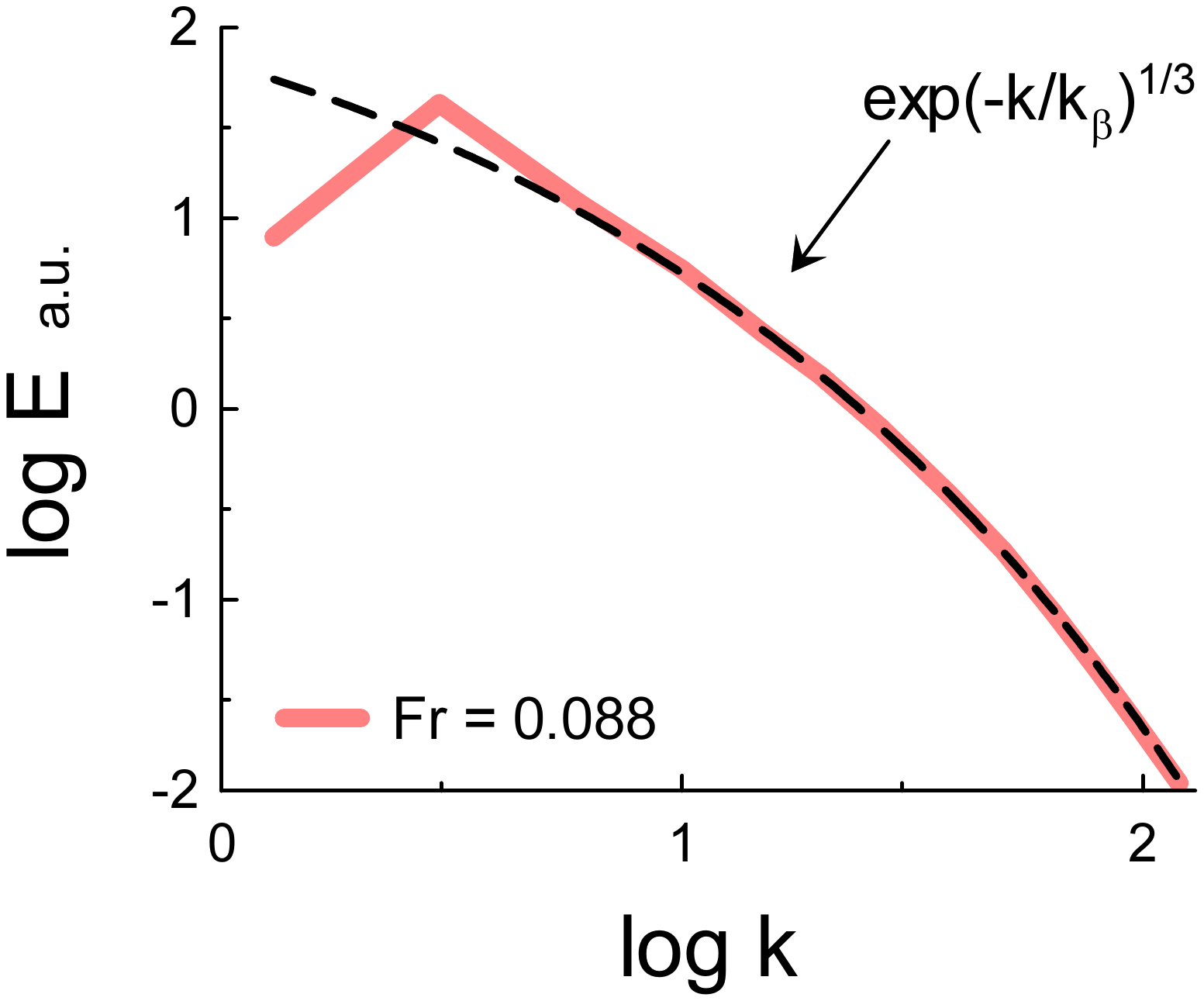}\vspace{-3.4cm}
\caption{\label{fig8} Total energy spectra ($k$ is isotropic wave number) for $Fr = 0.088$). The dashed line indicates the stretched exponential spectral law Eq. (5) with $\beta = 1/3$. }
\end{center}
\end{figure} 

 To complete the picture we show in figure 8 total energy spectrum of stably stratified turbulence without forcing and with the random isotropic initial conditions (a specific amount of
relative helicity was chosen at $t = 0$ in order to generate a helical flow). 
The data were taken from a recent Ref. \cite{ror} (Fig. 10a). The DNS was performed in the range $0.022 < Fr < 0.88$ ($Re = 6000$). All obtained power spectra are similar to that shown in the Fig. 8.  

\section{Unstably stratified turbulence in Rayleigh-Taylor mixing zone}

  In the recent Ref. \cite{bur} a DNS of an unstably stratified turbulence was performed in order to simulate development of turbulence in Rayleigh-Taylor mixing zone (see Introduction). Figure 9 shows a kinetic energy spectrum obtained in this DNS at the computing time $t = 3$ (Fr(t=0)=0.808, N=4, Pr = 1). The data were taken from Fig. 6 of the Ref. \cite{bur}. The dashed line indicates the stretched exponential spectral law Eq. (5) with $\beta = 2/3$ (Section II, subsection E). Figure 10 shows the spectrum at the computing time $t = 6$. The dotted arrow indicates tuning of the distributed chaos to the large-scale coherent structures.\\
  
    Figure 11 shows power spectrum of the vertical component of velocity field computed at time
$t = 3.1\tau$ in a recent DNS reported in Ref. \cite{bof} ($\tau =\sqrt{L_z/Ag}~$, $L_z$ is the vertical side of the computational domain). The data were taken from Fig. 9 of the Ref. \cite{bof}. In this DNS a more conventional model of the Rayleigh-Taylor mixing zone was used: without the term $sN{\bf u} \cdot {\bf e_z}$ in Eq. (2) but with the corresponding initial-boundary conditions (which can be transformed into the Eq. (2) by corresponding replacement of the variables, see Introduction). The dashed line indicates the stretched exponential spectral law Eq. (5) with $\beta = 3/5$ (Section II, subsection C). An indication of the inertial range is shown by the straight solid line with the slope -5/3 (cf Fig. 7). The dotted arrow indicates tuning of the distributed chaos to the large-scale coherent structures.\\

\begin{figure}
\begin{center}
\includegraphics[width=8cm \vspace{-1.1cm}]{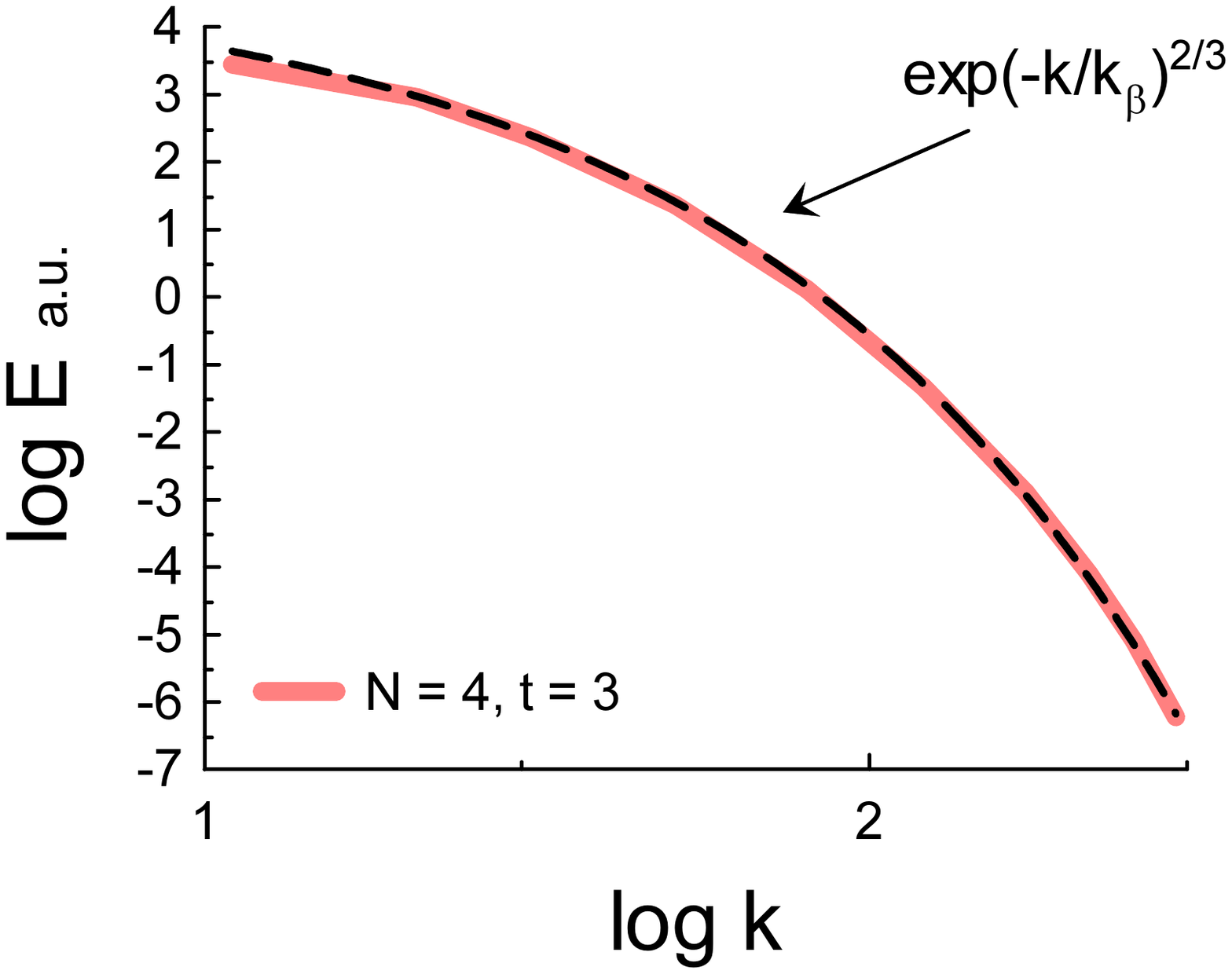}\vspace{-3.2cm}
\caption{\label{fig9} Kinetic energy spectra at the computing time $t=3$ ($Fr = 0.808, N=4$). The dashed line indicates the stretched exponential spectral law Eq. (5) with $\beta = 2/3$. }
\end{center}
\end{figure} 
\begin{figure}
\begin{center}
\includegraphics[width=8cm \vspace{-0.805cm}]{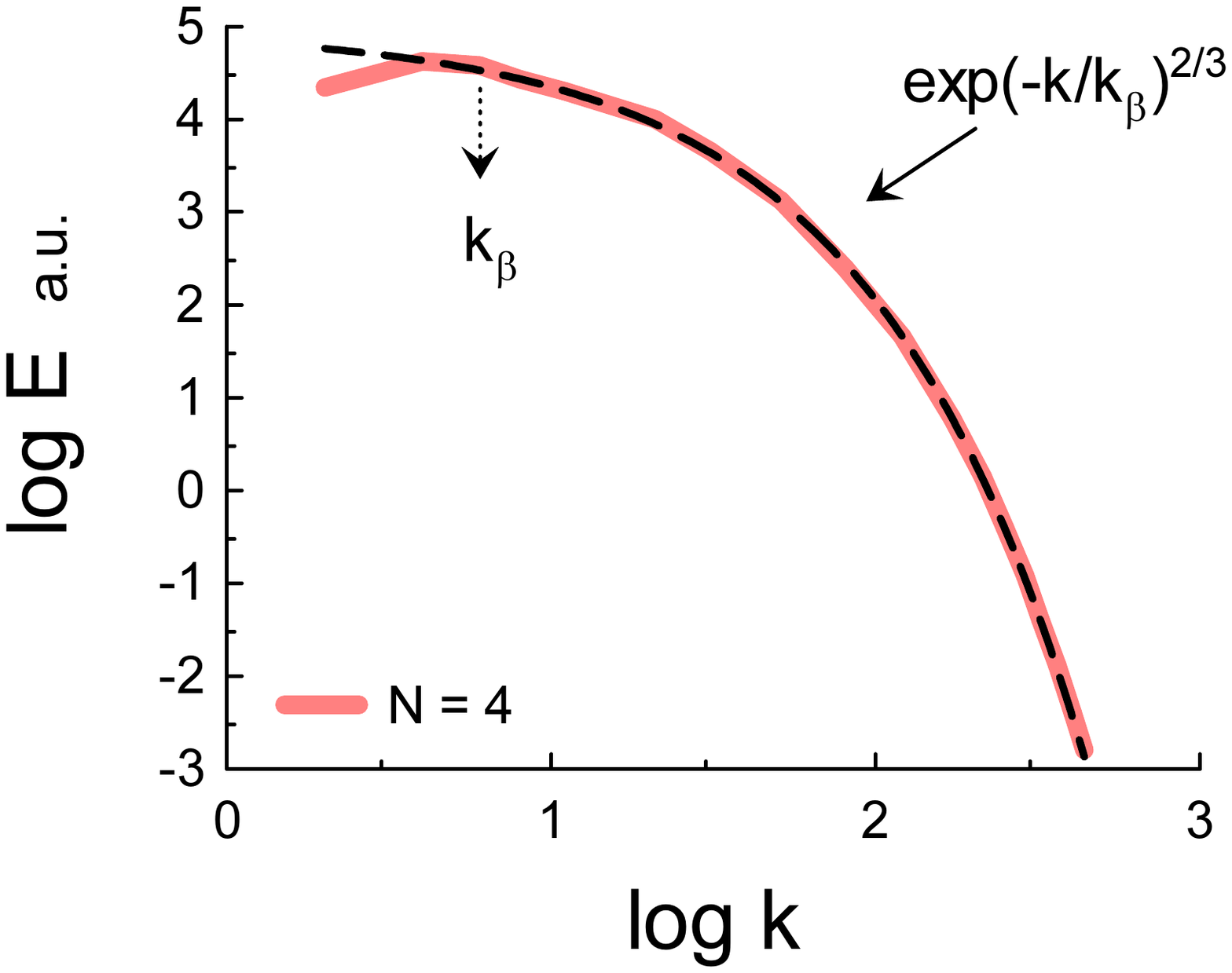}\vspace{-3.8cm}
\caption{\label{fig10} The same as in Fig. 9 but for $t=6$. }
\end{center}
\end{figure} 
\begin{figure}
\begin{center}
\includegraphics[width=8cm \vspace{-1cm}]{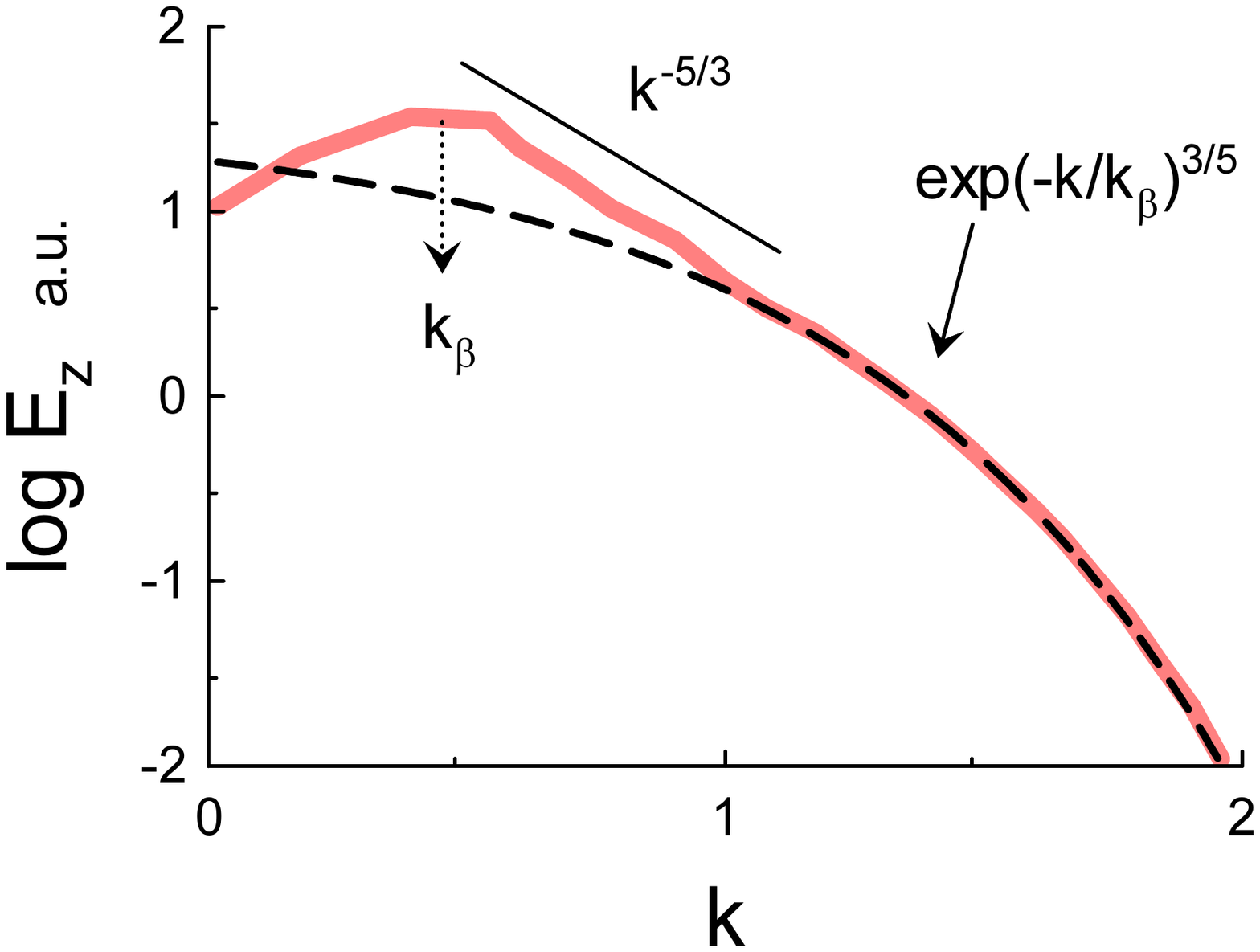}\vspace{-4cm}
\caption{\label{fig11} Power spectrum of the vertical component of velocity field. The dashed line indicates the stretched exponential spectral law Eq. (5) with $\beta = 3/5$.  }
\end{center}
\end{figure} 
\begin{figure}
\begin{center}
\includegraphics[width=8cm \vspace{-1.5cm}]{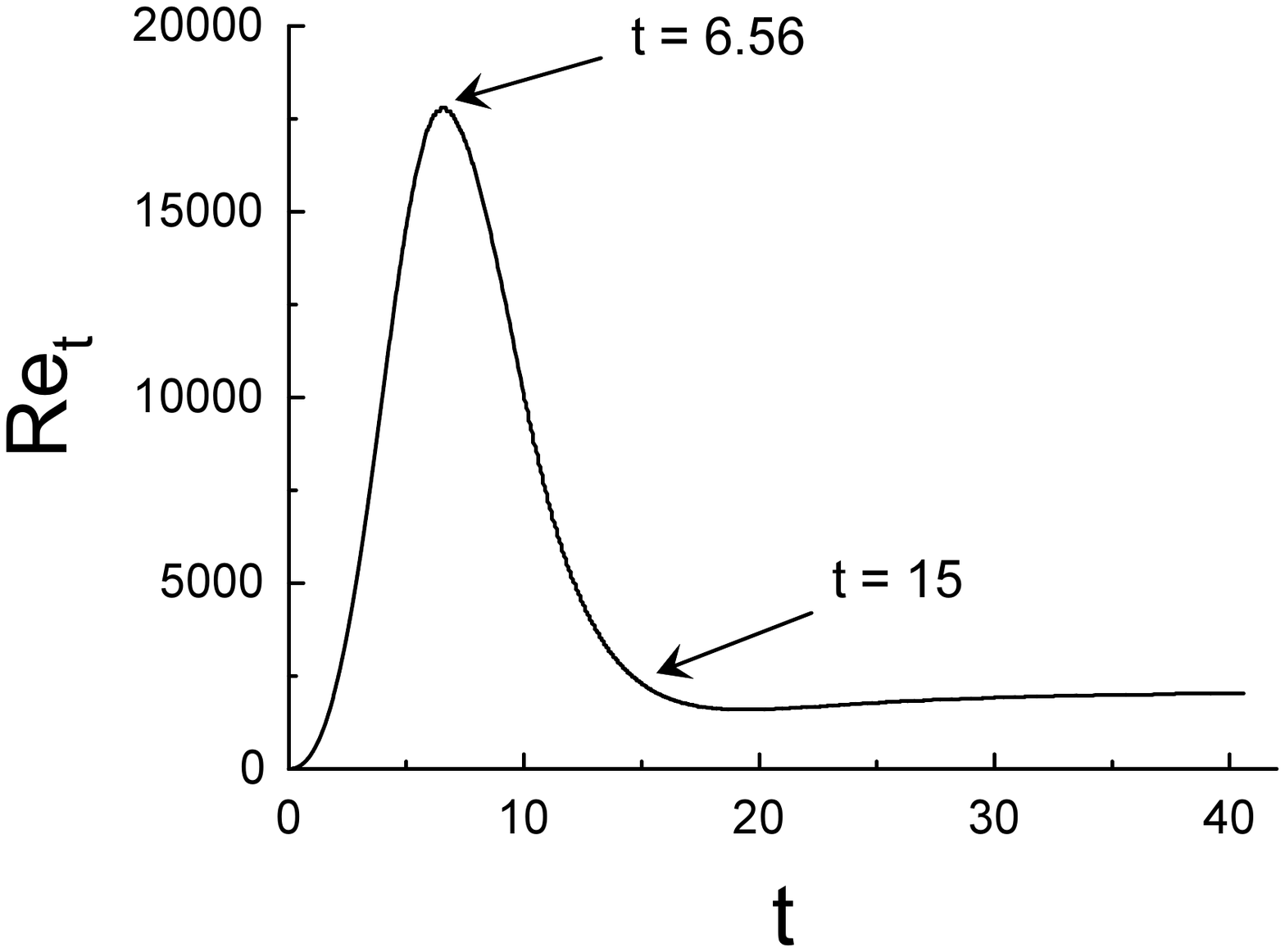}\vspace{-3cm}
\caption{\label{fig12}  Turbulent Reynolds number $Re_t$ against computing time.  }
\end{center}
\end{figure} 

   Another interesting model of the Rayleigh-Taylor mixing zone was used for a DNS in Ref. \cite{livescu} (see also Refs. \cite{livescu1},\cite{livescu2}). The equations used in this model are the two-fluid (with different molar masses) incompressible miscible Navier-
Stokes equations, which are obtained from the compressible Navier-Stokes equations in the limit $c \rightarrow \infty$  (c is the sound speed). The densities of the fluids remain constant. When the densities of the fluids are commensurate this approximation is a Boussinesq one. At this simulation $Fr = 1,~ A=0.05$. 

  Figure 12 shows turbulent Reynolds number $Re_t$ against computing time. Two crucial points are marked by arrows. Figure 13 shows 3D kinetic energy spectrum at $t= 6.56$. The dashed line indicates the stretched exponential spectral law Eq. (5) with $\beta = 3/5$ (Section II, subsection C). An indication of the inertial range is shown by the straight solid line with the slope -7/3 \cite{bb}. Figure 14 shows the 3D kinetic energy spectrum at $t= 15$. The dashed line indicates the stretched exponential spectral law Eq. (5) with $\beta = 2/3$ (Section II, subsection E). Analogous spectrum for $t = 40$ is shown in Fig. 15.
  
\section{Rayleigh-B\'{e}nard turbulent convection}

  Unlike the previous two cases the distributed chaos for the Rayleigh-B\'{e}nard turbulent convection in an upright cylindrical cell should be determined by the finite boundary effects: spontaneous breaking of the space translational symmetry by the finite boundaries conditions (Section II, subsection B: $\beta = 1/2$) and by the non-perfect orientation of the cell along the buoyancy direction (Section II, subsection D: $\beta = 4/7$). The only exclusion is the situation (related to the large scale circulation \cite{b3}) when an inertial range can deform the distributed chaos resulting in $\beta = 3/5$ (Section II, subsection C). The last case should be observed in spectral measurements performed in the large scale wind (near the side wall) \cite{b3}. 

\begin{figure}
\begin{center}
\includegraphics[width=8cm \vspace{-0.95cm}]{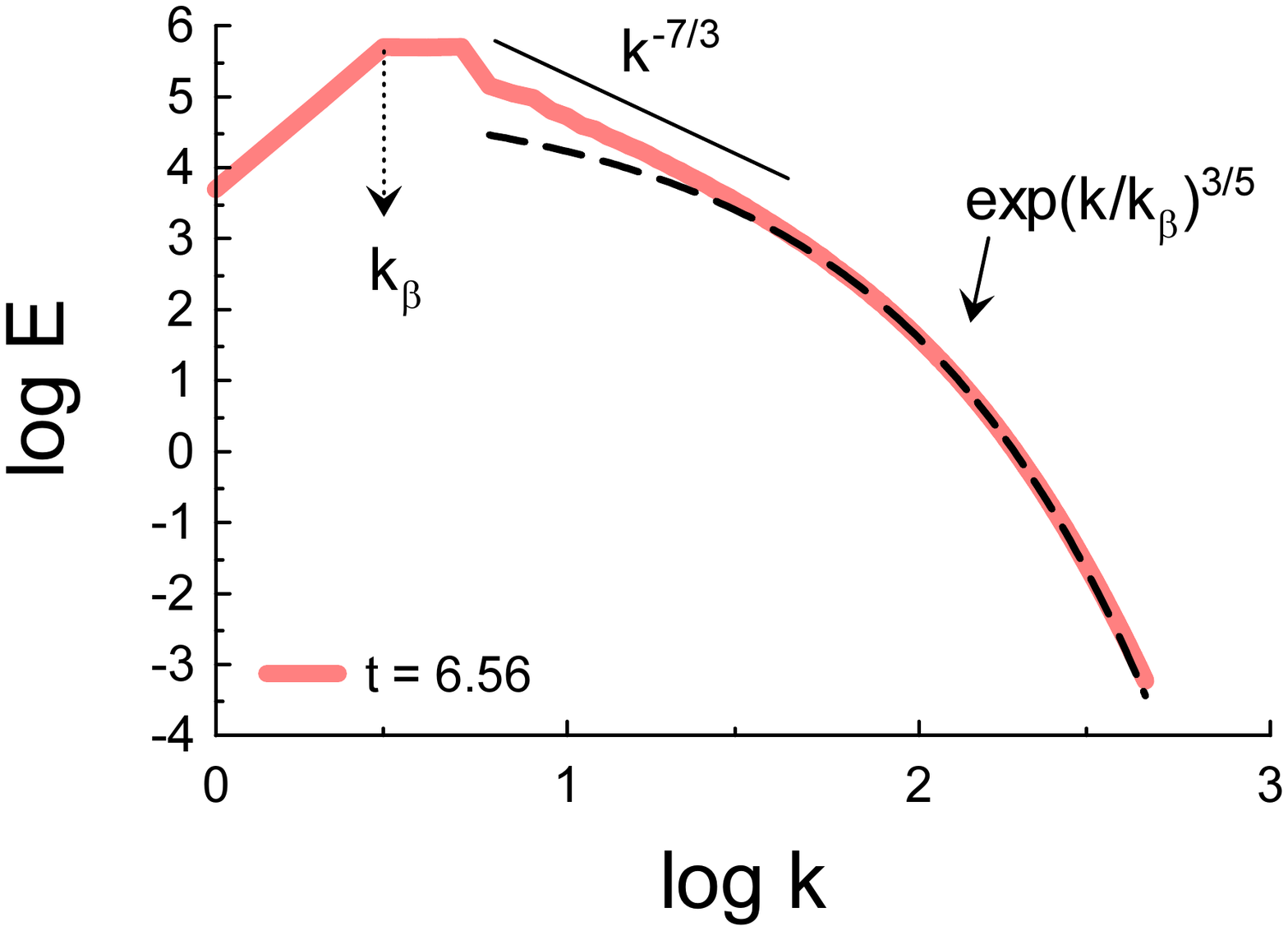}\vspace{-4.1cm}
\caption{\label{fig13} Kinetic energy spectrum at the computing time $6.56$. The dashed line indicates the stretched exponential spectral law Eq. (5) with $\beta = 3/5$. }
\end{center}
\end{figure} 

\begin{figure}
\begin{center}
\includegraphics[width=8cm \vspace{-1.5cm}]{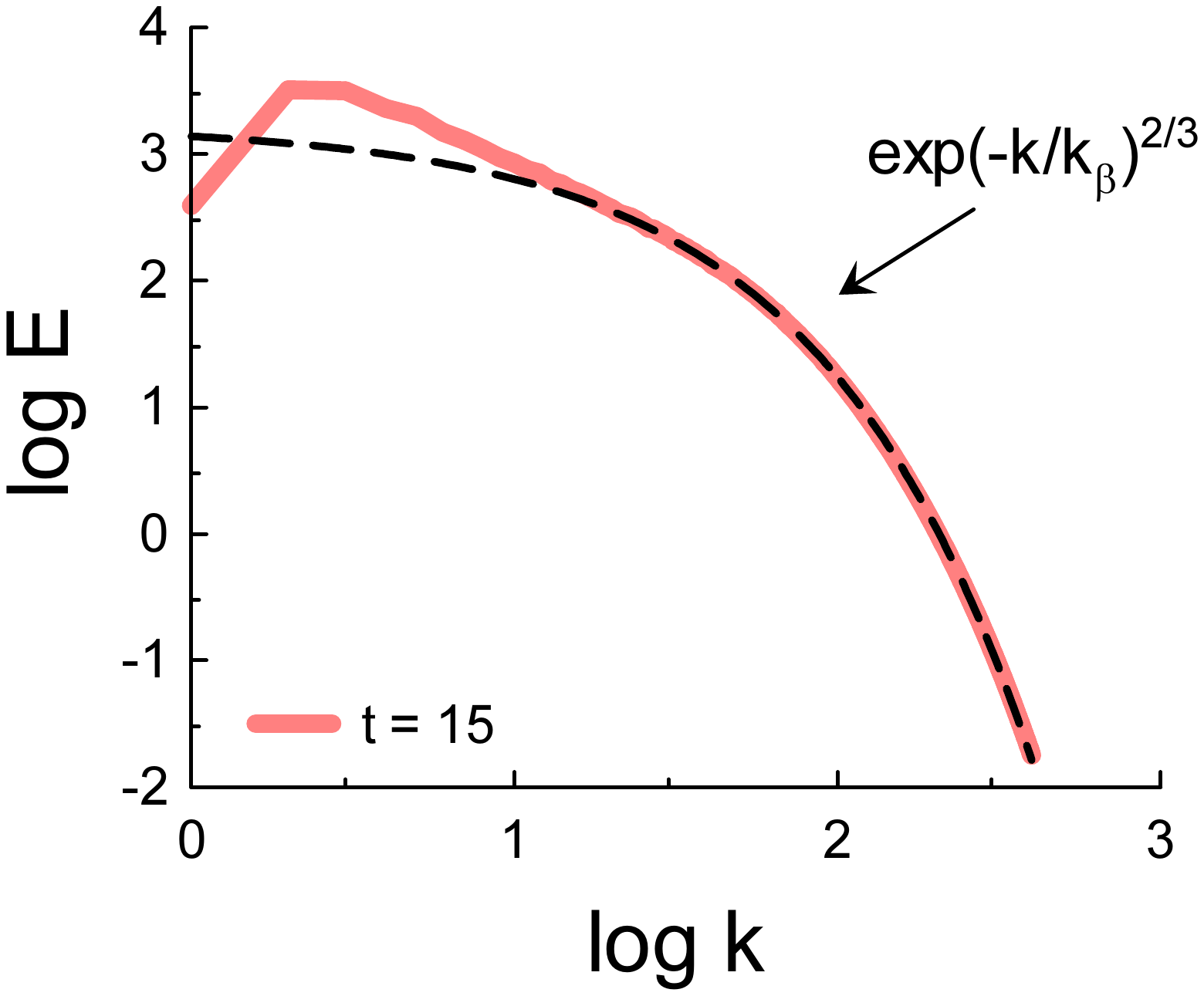}\vspace{-3.7cm}
\caption{\label{fig14} The same as in Fig. 13 but for the computing time $15$. The dashed line indicates the stretched exponential spectral law Eq. (5) with $\beta = 2/3$. }
\end{center}
\end{figure} 
\begin{figure}
\begin{center}
\includegraphics[width=8cm \vspace{-1.2cm}]{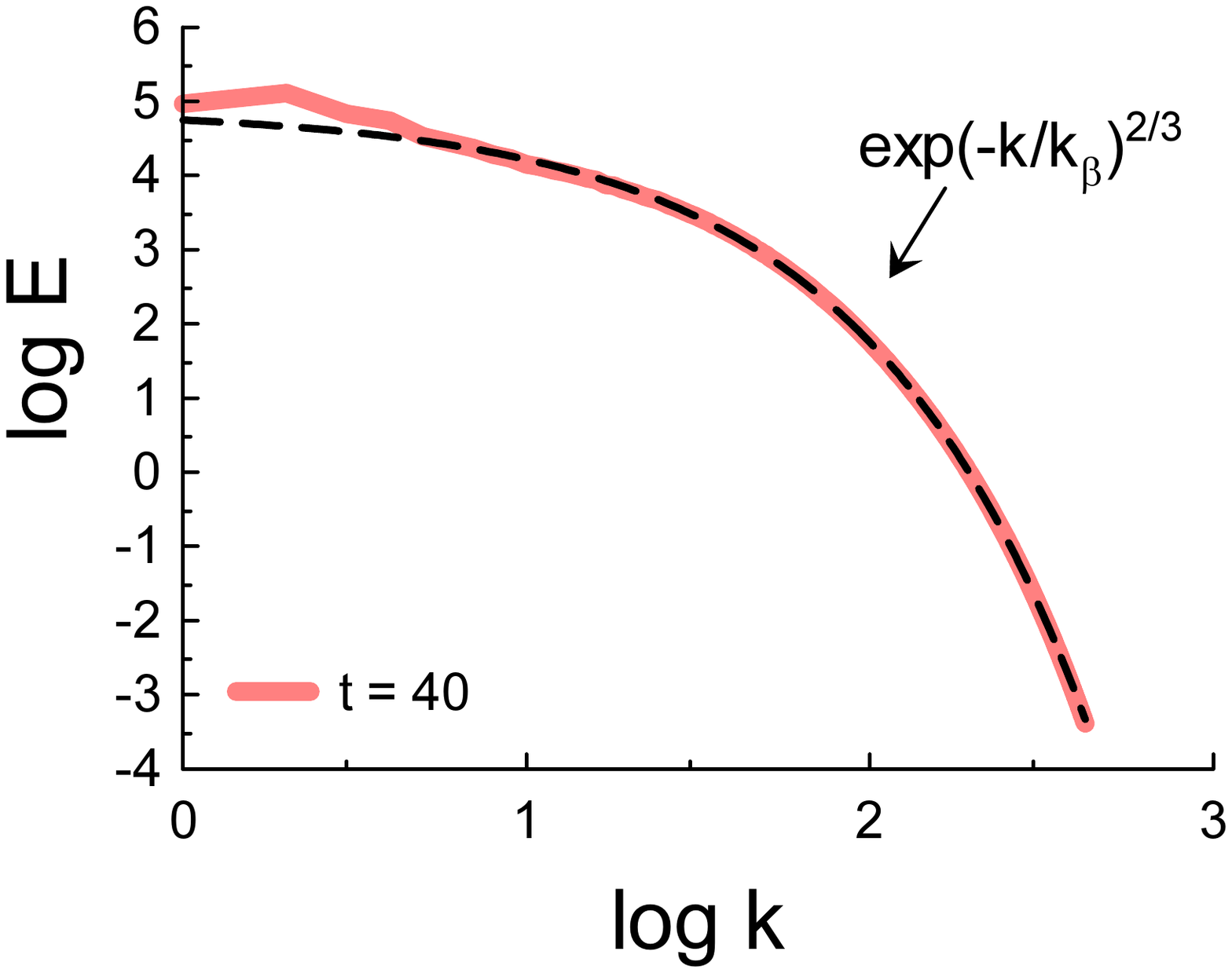}\vspace{-3.7cm}
\caption{\label{fig15} The same as in Fig. 14 but for the computing time $40$. The dashed line indicates the stretched exponential spectral law Eq. (5) with $\beta = 2/3$. }
\end{center}
\end{figure} 
  Figure 16 shows power spectrum of temperature for rather large Rayleigh number $3\cdot 10^{14}$ (Prandtl number $Pr=300$). The measurements were made at the center of an upright cylindrical cell \cite{as}. The authors of the Ref. \cite{as} noted that buoyancy completely suppresses inertial range in this experimental situation.  The dashed line is drawn in the Fig. 16  to indicate the stretched exponential spectrum with $\beta = 1/2$ (Section II, subsection B). The wavenumber spectrum Eq. (5) can be transformed into the frequency spectrum by means of the Taylor hypothesis \cite{my}.  
  
    Figure 17 shows power spectrum of temperature for Rayleigh number $4.6\cdot 10^{9}$. The measurements were made at the center of an upright cylindrical cell (with rough low and upper surfaces) \cite{du}. While the surface roughness results in the enhancement of the heat transport the temperature statistics in the center of the cell is approximately the same as for the smooth surfaces \cite{du}. However, the dashed line, drawn in the Fig. 17 in order to indicate the stretched exponential spectrum, shows $\beta = 4/7$ (cf Fig. 16). That indicates the spontaneous breaking of the space translational symmetry (homogeneity) by the non-perfect orientation of the cell along the buoyancy direction (${\bf g}$) in this experiment (Section II, subsection D).  
    
    It should be noted that for practical applications of the Rayleigh-B\'{e}nard turbulent convection just the case of the non-perfect orientation of the cell along the buoyancy direction is the most common one.

\section{Discussion}

\begin{figure}
\begin{center}
\includegraphics[width=8cm \vspace{-0.8cm}]{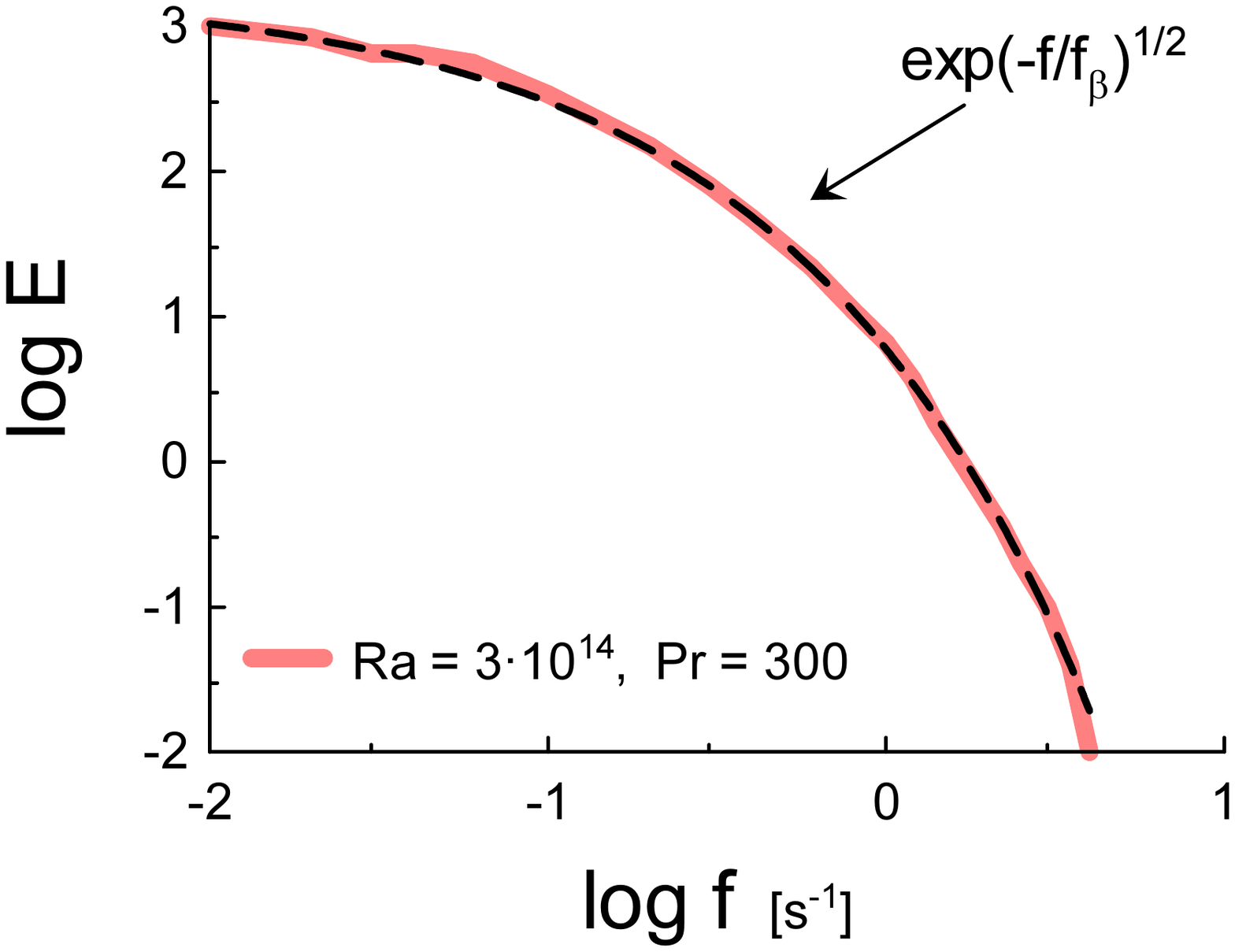}\vspace{-3.6cm}
\caption{\label{fig16} Power spectrum of temperature for rather large Rayleigh number $3\cdot 10^{14}$. The data were taken from Ref. \cite{as}. The dashed line indicates the stretched exponential spectral law Eq. (5) with $\beta = 1/2$. }
\end{center}
\end{figure} 

\begin{figure}
\begin{center}
\includegraphics[width=8cm \vspace{-0.9cm}]{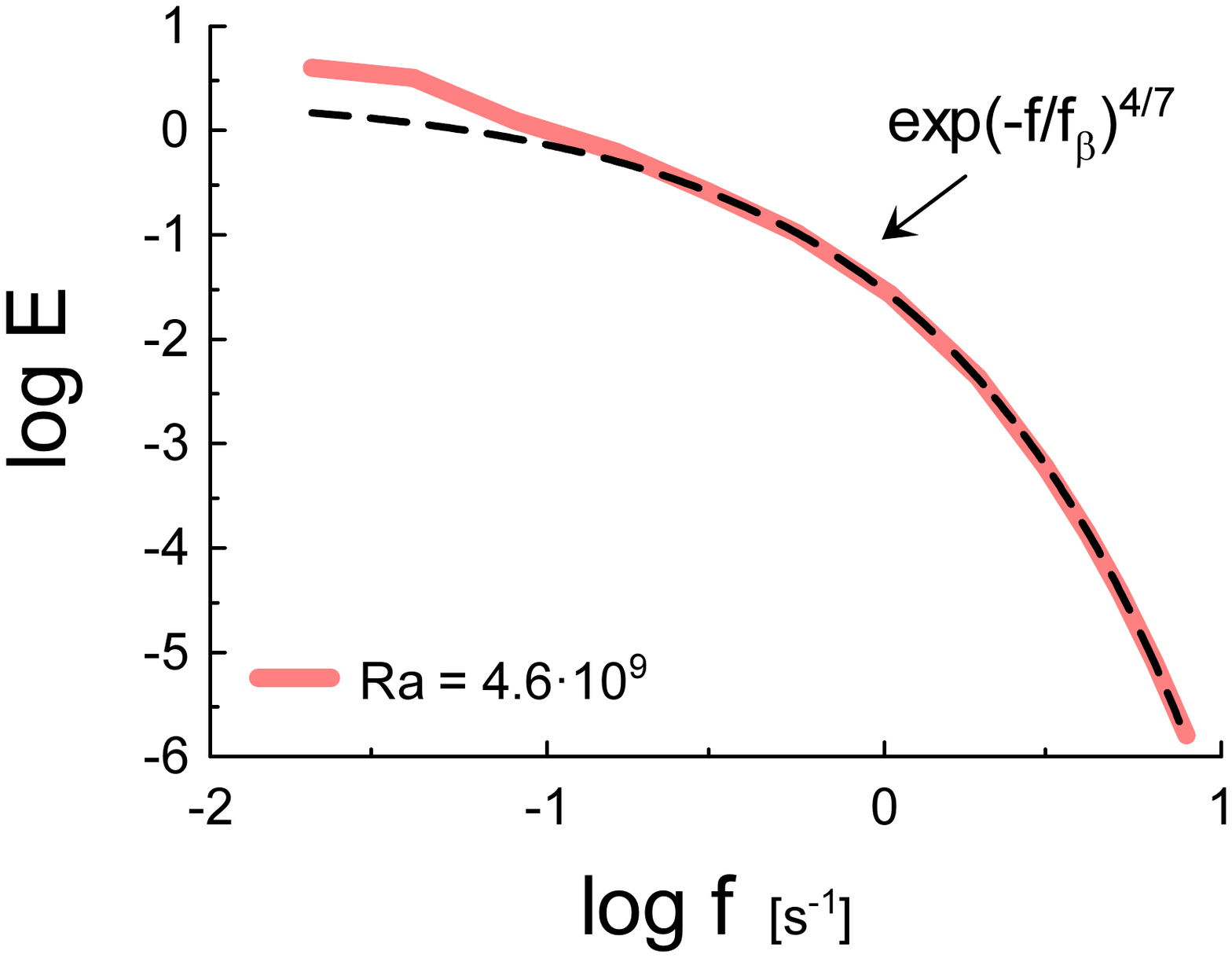}\vspace{-3.4cm}
\caption{\label{fig17} Power spectrum of temperature for Rayleigh number $4.6 \cdot 10^{9}$. The data were taken from Ref. \cite{du}. The dashed line indicates the stretched exponential spectral law Eq. (5) with $\beta = 4/7$. }
\end{center}
\end{figure} 
  It follows from the Figs. 1-8 that the distributed chaos with $\beta = 1/3$  (i.e. determined by the helicity correlation integral Eq. (17)) is the most common feature of the stably stratified turbulence. It is not surprising taking into account the crucial role of the waves in this type of the buoyancy driven turbulence. It is also clear that these waves mostly dominate spectral properties of the vertical component of velocity field, while the horizontal component is dominated by the diffusive processes both for the weak and strong stable stratification ($\beta =2/3$ (Section II, subsection E). For the last case influence of the low boundary can overcome the wave effects and result in $\beta =1/2$ for the vertical component of the velocity field (the spontaneous breaking of the space translational symmetry: Section II, subsection B). Appearance of an inertial range of scales (even an embryonic one) results in deformation of the distributed chaos and in $\beta = 3/5$ - Fig. 7 (Section II, subsection C). This is a common feature of all the types of turbulence (including isotropic and homogeneous one) \cite{b3}. \\
  
  For the unstably stratified turbulence in the Rayleigh-Taylor mixing zone Figs. 9-15 indicate the diffusive (viscosity) processes - $\beta =2/3$, as the most common dominating ones in the anisotropic chaotic mixing of the two fluids under buoyancy forces. The inertial processes ($\beta =3/5$) overcome this effect for sufficiently strong turbulence only Figs. 11-13. \\
  
  The distributed chaos in Rayleigh-B\'{e}nard turbulent convection in an upright cell is determined by the strong confinement conditions. That is: the spontaneous breaking of the space translational symmetry (homogeneity) by the finite boundaries Fig. 16 (Section II, subsection B) or by the non-perfect orientation of the cell along the buoyancy direction Fig. 17 (Section II, subsection D). About the effects of the large-scale circulation (thermal wind) on the distributed chaos near the side wall see Ref. \cite{b3}.

\section{Acknowledgement}

I thank D. Livescu for sharing his data.


\begin{thebibliography}{99}

\bibitem{bur}A. Burlot et al., J. Fluid Mech. {\bf 765}, 17 (2015).
\bibitem{ll2} L.D. Landau and E.M. Lifshitz, Mechanics (Pergamon Press 1969).
\bibitem{saf} P. G. Saffman, J. Fluid. Mech. {\bf 27}, 551 (1967).
\bibitem{dav1} P. A. Davidson, J. Phys.: Conference Series {\bf 318} 072025 (2011).
\bibitem{dav2} P. A. Davidson P.A. Turbulence in rotating, stratified and electrically conducting fluids. (Cambridge University Press, 2013).
\bibitem{b1} A. Bershadskii, arXiv:1512.08837 (2015).
\bibitem{nrz} A. C. Newell, B. Rumpf, V. E. Zakharov, Phys. Rev, Lett., {\bf 108},
194502 (2012)
\bibitem{b2} A. Bershadskii, arXiv:1601.07364 (2016).
\bibitem{my} A. S. Monin, A. M. Yaglom, Statistical Fluid Mechanics, Vol. II: Mechanics of Turbulence (Dover Pub. NY, 2007).  
\bibitem{b3} A. Bershadskii, arXiv:1604.07762 (2016).
\bibitem{lt} E. Levich and A. Tsinober, Phys. Lett. A {\bf 93}, 293 (1983).
\bibitem{l} E. Levich, Concepts of Physics {\bf VI}, 239 (2009).
\bibitem{b4} A. Bershadskii, arXiv:1604.05211 (2016).
\bibitem{katul} D. Li, G. G. Katul, and E. Bou-Zeid, Boundary-Layer Meteorol., {\bf 157}, 1 (2015).
\bibitem{rem} S. Remmler, S. Hickel, Theor. Comput. Fluid Dyn., {\bf 27}, 319 (2013).
\bibitem{rmp} C. Rorai, P.D. Mininni and A. Pouquet, Phys. Rev. E {\bf 92}, 013003 (2015).
\bibitem{ror} C. Rorai, D. Rosenberg, A. Pouquet, and P. D. Mininni, Phys. Rev. E {\bf 87}, 063007 (2013).
\bibitem{bof} G. Boffetta, A. Mazzino, and S. Musacchio, Phys. Rev. 83, 056318 (2011).
\bibitem{livescu} http://turbulence.pha.jhu.edu/datasets.aspx (Section 4).
\bibitem{livescu1} D. Livescu, Phil. Trans. R. Soc. A {\bf 371}, 20120185 (2013).
\bibitem{livescu2} D. Livescu et al., J. Turbul. {\bf 10}, N13 (2009).
\bibitem{bb} A. Bershadskii and G. Branover, J. Phys. I, {\bf 4}, 1115 (1994).
\bibitem{as} S. Ashkenazi and V. Steinberg, Phys. Rev. Lett. {\bf 83}, 3641 (1999).
\bibitem{du} Y.-B. Du and P. Tong, Phys. Rev. E {\bf 63}, 046303 (2001).



\end{thebibliography}
\end{document}